\documentclass[a4paper]{article}
\def\be{\begin{equation}}
\def\ee{\end{equation}}
\def\ct{\cite}
\def\bi{\bibitem}

\begin{document}
\date{February 5, 2001}

\title{Vacuum charges within a teleparallel weyl tensor:  a new apporach to quantum gravity}
\author{Edward F. Halerewicz, Jr.\footnote{email:  ehj@warpnet.net} \\ \\ Lincoln Land Community College\footnote{This address is given for mailing purposes only, since I'm  a student and do not hold a professional position at the above address.  This work was made possible through my own personal research and studies, and does not reflect the above listed institution.}
\\ 5250 Shepherd Road\\ Springfield, IL 62794-9256 USA\\}
\maketitle

\begin{abstract}
A comparison is given between the Newtonian and Einsteinian frames
of gravitation.  From this it is shown that there exist a weak
connection to gravitation and electromagnetism.  This connection
is then studied more thoroughly with the Weyl tensor and with the
electromagnetic vacuum $\Lambda$.  Which dictates General
Relativity should be reformulated to confer to a
`Einstein-Cartan-Weyl' geometry.  Where it is seen that the
Gravitational Constant is the inverse of the Compton wavelength
shown through a Weyl gauge potential of form
$[F_{\alpha\beta}+A_{a\beta}^\alpha]_{;\beta}$.  The gauge
potential along with Einstein-Cartan geometry is argued to explain
superluminal velocities observed within General Relativity.
\end{abstract}

{\flushright PACS numbers:  04.20.-q, 95.30.Sf, 98.80.Es, 04.60.-m, 06.20.Jr}

\begin{flushleft}
``{\it The acceleration of motion is ever proportional to the
motive force impressed; and is in the direction of the right line
in which that force is impressed.}" \flushright --Newton (1687)
\end{flushleft}
\maketitle

\section{Introduction}
There are currently two viable models for the gravitational field, being the classical and the relativistic gravitational field.  Of course this refers to the law of universal gravitation proposed by Newton and General Relativity (GR) by Einstein.  The two theories are certainly correlated and attempt to describe the same phenomena, however mathematically they are treated quite different from one another. It thus becomes natural for the sake of coherence to give a unified structure of the two theories, even if only ad hock.  What is peculiar, or even far more obvious is that both theories of gravitation are subtle variations of the {\em second law of motion}.

The development of this work was made possible through studying
the relationship between classical and relativistic gravitational
fields where a weak connection can be shown to electromagnetism.
In previous works it has been shown that the cosmological constant
$\Lambda$ can be represented by a covariant electromagnetic field
\ct{ve2}, \ct{ve3}.  It has also been suggested that the
cosmological constant can be derived from the quantum vacuum
\ct{ccv}, \ct{ve}.  Using an analog of quantum theory and
electromagnetism an empirical unification with gravitation is
quickly realized with the classical Kaluza-Klien (KK) theory
\ct{KK}.  From an empirical KK geometry a connection with
gravitation and Newton's {\em second law of motion} can be
explained by means of the Weyl tensor within the GR formalism.
The general conclusion that can be arrived from this analysis is
that GR is not a complete theory of gravitation when only
considering Ricci curvature of the Riemannian manifold.

The organization of this work is given rather straight forward in
\S \ref{neg} the Newtonian theory for gravitation is explained
through the {\em second law of motion}.  In \S \ref{ast} Newton's
view of the world is briefly discussed, where as in \S \ref{sr}
Einstein's `world view' is given, where the two are related within
\S \ref{ftf}.  In \S \ref{dag} G is reformulated through the
second law, where a relation is shown to electrostatics in \S
\ref{elecs}, which is shown to be a superposition field within \S
\ref{sp}.  In \S \ref{emper} empirical equations of this field are
given, which is resembles a KK space which is discussed in \S
\ref{kks}.  A gauge field is considered for gravitation in \S
\ref{gback}, which suggest a correlation to the vacuum energy
explored in \S \ref{veg}.  In \S \ref{emwt} the GR analog is
discussed with the Weyl tensor.  In \S \ref{glag} first order
Lagrangians are presented, which leads to teleparallel Weyl tensor
in \S \ref{tpg}.  In \S \ref{vmg} a relationship between vacuum
energy and the gravitational constant are given.  A standing wave
is shown in \S \ref{swf} which gives an allusion of the classical
KK space.  The explanation for the Vacuum charges in the title is
seen in the Appendices.  Appendix \ref{comp} describes the
relation between the gravitational constant and the Compton
wavelength, as well as explaining superluminal velocities observed
in astrophysics.  Appendix \ref{lmass} gives an alternative origin
for mass increase, Appendix \ref{gqd} briefly discusses other
theoretical values for G.  Finally in Appendix \ref{impl} there is
shown a need to modify the definition of the planck length under
this work.

\subsection{Newton-Einstein Gravity\label{neg}}
It is obvious to start such a modeling with the widely known
Newton-Einstein action:
\be
{d^2x_{\mu}\over dt^2}={\kappa c^2\over 8\pi}{\partial\over\partial x_{\mu}}\int {\sigma d V_0\over t}
\ee
Where the left term is defined by the second law of motion:
\be
m\left\{ {d^2 x^{\mu}\over dt^2}+\Gamma^{\alpha}_{\beta
y}{dx^{\beta}\over dt}{dx^y\over dt}\right\} =F^{\alpha} \ee Here
another generalization of the second law can be given with the equation
$\vec F=m{d^2\vec r\over dt^2}$.  When taking the convection $\vec
r=(x,y,z)$ a gravitational acceleration is derived by:
\be
m{d^2 \vec r\over dt^2}=-{GMm\over\left| {\vec r -\vec R'}\right|
^2}\:\vec e(\vec r,\vec R) \ee The gravitational field is then
defined through $\varphi =GM/R$, thus a gravitational field is
produced through poisson's equation through $\nabla\varphi =4\pi
G_{\rho}$.  Where for simplicity sake we receive the standard
deviation for Newton's gravitational field:
\be
\vec F_g=G{m_1m_2\over d\vec r^2} \ee It is quite clear that
Newton's formulation of gravitation is formed through the his
second law of motion.  Which is explained as an external force
mechanism which causes masses to accelerate one another.

\subsection{the absolute spacetime\label{ast}}
Newton considered space and time as separate and finite invariant
dimensions.  We can see this definition early on in Book I of {\it
Princpa} by means of Scholium I:
\begin{quote}
``{\it absolute, true, and mathematical time, of itself, and from
its own nature flows equably without regard to anything external,
and by another name is called duration:  relative, apparent, and
common time is some sensible and external (whether accurate or
unequalable) measure of duration by the means of motion, which is
commonly used instead of true time; such as an hour, a day, a
month, a year}."
\end{quote}
A conclusion that is drawn from the roots of Euclidean geometry
which can be expressed in the form
\be
ds=\sqrt{dx^2+dy^2+dz^2} \ee where the following spatial identities arise by
means of an infinitesimal rotation:
\be
F^x=F^xcos\theta +F^ysin\theta ,\quad F^y=F^xsin\theta
+F^ycos\theta ,\quad F^{z'}=F^z \ee again we see how the spatial
definition directly relates to the second law of motion.

\subsection{Special Relativity\label{sr}}
With a suggestion from Minkowski Einstein transformed the
cherished absolute description of space and time to a relative
space-time of the form:
\be
ds^2=dct^2-dx^2-dy^2-dz^2 \ee with the invention of a `spacetime'
continuum, one can notice subtle changes with an infinitesimal
rotation:
\be
ct'=cosh\theta ct-sinh\theta x,\quad x'=sinh\theta ct+coshx,\quad
y'=y,\quad z'=z \ee With Einstein's fundamental postulate
acceleration in such a frame would be limited to the speed of
light, lending the beta function:
\be
\gamma =\gamma (v)={1\over\left( {1-{v^2\over c^2}}\right)^{1/2}}
\ee working backwards now we can see that kinetic energy of a body
in this frame is given in the form
\be
KE=\int\limits_0^u m\gamma  ^3udu=mc^2(\gamma -1) \ee Thus at this
point we see that these two different formulations of space will
produce very different forms of acceleration.  In classical terms
mass is defined as a focused point of force, while in relativistic
terms it is defined as `stress-energy' within an arbitrary
manifold.

\subsection{forces to fields\label{ftf}}
Taking a new look at Newton-Einstein gravity one may make a second
order generalization of the field by:
\be
T^i_j\left\{ {d^2x_{\mu}\over dt^2}-\Gamma^{\alpha}_{\beta
y}{dx^{\beta}\over dt}{dx^y\over dt}\right\} ={kc^2\over
8\pi}g^i_j{\partial\over\partial x_{\mu}} \ee Notice how this
equation describes gravitation not as a force but as a manifold.
Furthermore mass is no longer an intrinsic property but a local
field in the geometry, i.e. the change results in going from a
point particle theory (mass) to a field theory (tensors).  This
force is taken equivalent to a manifold of the form ${\partial
f\over \partial x^{\mu}}={\partial\over\partial x^{\mu}}$,
assuming a Riemannian manifold and lines of calculations one results in the
Einstein-Field-Equation (EFE)
\be
R^i_j-{1\over 2}g^i_jR=-{kc^2\over 8\pi}T^i_J. \label{efe} \ee
Thus it can naturally be seen that the roots of GR
can be originated through the second law of motion.  In a larger
since, the ``gravitational force" is in reality a consequence of the second law of motion, expressed differently only in the terms of mathematical
dimensions.  Therefore one may wish to explain the gravitational
force as a two dimensional acceleration of the form
$F'_g=a(E/c^2)(E/c^2)/dr^2 =4\pi a_{\rho}=\nabla\varphi$.  Of
course for a gravitational field a is replaced by Newton's
Gravitational constant G.

\subsection{dimensional analysis of the gravitational constant\label{dag}}
One can not deny the similarity between a classical gravitational
field and the Coloumb Law $F_{col}=kQ_1Q_2/r^2$. Suggesting
classically what Kaluza, Klein, Weyl, and others have proposed, a
unification with the electromagnetic force.  In GR the flat or
massless gravitational field is given with $R^i_j-{1\over
2}g^i_jR=0.$ This is not entirely correct because the kappa term
does not entirely vanish leading to $R^i_j-{1\over 2}g^i_jR=8\pi
k_e\dots$. It is seen that the interpretation of a geometrical
manifold neglects gravitational acceleration.  If it is a property
of the electromagnetic field however, the second law of motion and
the vacuum field equations still hold true.

To elaborate more on the gravitational constant\footnote{Modern values given the gravitational constant as \ct{G}:
$G=(6.74215\pm 0.000092)\times 10^{-11}m^3 kg^{-1}s^{-2}$.}  one must be familiar with its roots, where
one begins with Kepler's third law of motion:
\be
T=2\pi\sqrt{{a^3\over M}}. \ee In keeping with the relationship between the gravitation and the second law of motion this must be rewritten in the form $T^2=kr^3$, which is analogous to pendulum
motion $T^2=4\pi ^2\:l/g$ where
\be
k\equiv\sqrt{G}={2\pi\over T}\sqrt{a^3\over M}. \ee Through
dimensional analysis we can reduce this in a form which relates to
the Coloumb law:
\begin{eqnarray}
F=m\times{(2\pi r/T)^2\over r}=m\times {4\pi ^2mr\over T^2}=m\times {(2\pi r)^4/kr^3\over r}\\=m\times {16\pi r/k\over r}=m\times {16\pi mr\over kr^3}=m\times {16\pi m\over kr^2}=G.\nonumber
\end{eqnarray}
Once again we see the relevance of the second law of motion,
perhaps more relevently with Kepler's Laws.  Furthermore the gravitational constant G can be represented by $k\equiv\sqrt{kr^2}$ such that Newton's Law of universal gravitation becomes:
\be
F=\sqrt{km_1m_2r}
\ee

\subsection{electrostatics\label{elecs}}
In more simpler language the force of gravitation can be derived
through the gaussian gravitational constant $k$ of a line charge
by means of a Coloumb field.
\be
\delta \oint Lds_{col}=(kQ_1Q_2\vec r)^{1/2} \ee Where an electric
field is propagated perpendicularly by:
\be
E_{\perp}=\lambda s\int\limits_{-L/2}^{+L/2}(z^2+s^2)^{-3/2}dz=\lambda s {1
\over s^2}\frac{L}{{\left( {\frac{{L^2 }}{4} + s^2 } \right)^{1/2} }}
\ee
or simply
\be
E_{\perp}={2\lambda L\over s}(L^2=4s^2)^{-1/2} \ee with poisson's
equation a general electrostatic potential is given by $\nabla
^2\phi=-4\pi\rho (\vec r)$ whence by the fundamental theorem of
vector fields we have an inverse square relationship
\be
\phi=\oint dV{\rho\over R}={\rho\over R}dxdydz=\int{\lambda
dz\over R} \ee for simplicity we will look at a charge
configuration of the form $E_r={\partial\phi\over\partial r}\vec
r$.  We now notice a direct relationship between an electrostatic
field line and gravitational acceleration by $\vec
g={\partial\varphi\over\partial r}\hat r$.  Empirically the
combination of the two fields would represent a force of the
first order
\be
\vec F_g={\partial\phi\over\partial r}\vec
r+{\partial\varphi\over\partial r}\hat r=\sum{\partial ^2\phi
^2\over\partial ^2 r^2}\hat r^2=\int^r_{\infty}{Gm_1+m_2\; r\over
r^2}=\sqrt{k\partial Q_1\partial Q_2 (\vec r).} \ee From here it
can bee seen that a (neutral) static charge configuration can yield gravitational acceleration.
\be
g_{ij}=-{\partial ^2\phi ^2\over\partial ^2 x_{ij}}=-\phi _{ij}
\label{ga} \ee Such that it is now seen that relative acceleration
of two particles can be given in pseudo Levi-Civta
coordinates
\be
{d^4x^{ij}\over dt^4}=\Delta g_{ij}=-\phi _{ijkl}\eta ^{kl}. \ee
Where a generalized pseudo Riemannian field is produced
\be
R^*_{abcd}-{1\over 2}g_{cd}\:^{pq}R_{pgra}=-8\pi GT_{abcd} \ee
which reduces to directly to the Einstein Field Equation (\ref{efe}).

\section{superposition}\label{sp}
Equation (\ref{ga}) can be represented by an operator of the form
$i\hbar$ such that
\be
-\phi _{ij}={h\over i}{\partial ^2\phi ^2\over\partial ^2 x_{ij}}+H\psi
\label{hop}
\ee
with Schr\" odinger's equation one has
\be
i\hbar={\partial\psi\over\partial t}=-{h^2\over 2m}\left(
{\partial ^2 \phi\over\partial x_{ij}}+{\partial
^2\varphi\over\partial x_{ij}}\right) \label{qf} \ee from the
Laplacian $\nabla ^2 a$ we note this represents the original
field, and which yields two gradients in spherical coordinates of
the form $$ \vec\nabla a={\partial a\over r}\hat r+{1\over
r}{\partial a\over\partial\theta}\hat\theta +{1\over r\:
sin\theta}{\partial a\over\partial\phi}\hat\phi $$ Which gives
rise to electrostatic configurations and gravitational
acceleration.  Which naturally lends itself to the Schwartzschild
solution when the fields are given in the first order
approximation in the classical field
\be
ds^2=(1-2\varphi )dt^2-{dr^2\over (1-2\varphi )}-r^2d\theta^2-r^2sin^2\theta d\phi ^2.
\ee

The gravitational and electrical fields in equation (\ref{qf}) can
be related more clearly through superposition.  This also means
that the field equations (\ref{ga}-\ref{qf}) are really
superposition fields.

A superposition of electric and gravitational fields can be given
through $\psi (x)=\psi _{\phi}(x_s)+\psi _{\varphi}(x_s)$, with
Huygens principle yields:
\be
\psi (x)\sim \int _{\phi}{\exp[2\pi i(x-x_s)/\lambda]\over\left|
x-x_s\right|}\psi _{\phi}(x_s)dx_s+\int _{\varphi}{\exp[2\pi
i(x-x_s)/\lambda]\over\left| x-x_s\right|}\label{sup}\psi
_{\varphi}(x_s)dx_s. \ee Where through quantum mechanics an
interference between the two fields arises from the probability
\be
P(x)=\left| \psi _{\phi}(x)\psi _{\varphi}(x)\right|
^2=P_{\phi}(x)+P_{\varphi}(x)+\psi ^* _{\phi}(x)\psi
_{\varphi}(x)+\psi ^* _{\varphi} (x)\psi _{\phi}(x) \ee thus
equation (\ref{ga}) may be reevaluated in the form:
\be
g_i=-{\partial\psi (x)\over\partial x_i}=-\psi (x)_i
\ee
lending
\be
{d^2x^i\over dt^2}=-\Delta g_i=-\psi (x)_{ij}\eta ^j
\ee

\subsection{empirical equations\label{emper}}
\label{empe} From the above the empirical gravitational field that
translates is
\be
R^{\mu\nu}-{1\over 2}\delta ^{\mu\nu}R={8\pi G\over c^4}T_M^{\mu\nu}+{8\pi k_e\over c^4}{a\over m}T_{CC}^{\mu\nu}
\ee
or
\be
R^{\mu\nu}-{1\over 2}\delta ^{\mu\nu}R={8\pi G\over c^4}T_{\psi
(x)}^{\mu\nu} \ee Since this field describes a quantum
superposition, imaginary coordinates are required lending:
\be
^*R^*_{abcd}-{1\over 4}\epsilon_{ab}\:^{pg}\epsilon_{cd}\:^{ra}R_{pgra}=i\left\{ {8\pi G\over c^4}[T_M^{abcd}+T_{EM}^{abcd}(Q_1)+T_{EM}^{abcd}(Q_2)]\right\}\hbar
\ee
Of course this would correspond to a complex spacetime
\be
\phi (x,y,z,t)=\int_{-\pi}^{\pi}F(xcos\theta +ysin\theta
+iz,y+iz\:sin\theta+cos\theta,\theta)d\theta \ee Maintaining the
Minkowski metric, the background manifold $\cal M$ one has
\be
(\omega, z^2)=\omega
ct^2-(\omega)z^2_{1}-(\omega)z^2_{2}-(\omega)z^2_{3} \label{com}
\ee Without the superposition of the mass-energy tensor, the
vacuum field equation becomes:
\be
R_{\mu}^{\nu}-{1\over 2}g_{\mu}^{\nu}R={8\pi k_e\over
c^2}T_{EM}^{\mu\nu}(Q_1)+T_{EM}^{\mu\nu}(Q_2). \ee From equation
(\ref{hop}) it is seen that a quantum interpretation must be given to
G.  With electrodynamics in mind one might consider a form which
pertains to the fine structure constant
\be
\alpha_e={2\pi e^2\over\hbar c}\rightarrow\alpha_g=-{1\over
2}{4\pi Gm^2\over\hbar c}.\label{fine} \ee This interpretation can
be made when one takes the Weyl tensor, and compares it to the
mechanical properties of an electromagnetic field:
\be
{\partial T_i^{\alpha}\over\partial k}-{1\over 2}{\partial
g_{\alpha\beta}\over\partial x_i}T^{\alpha\beta}=0. \ee As
suggested in the beginning of this work the above field is
implicitly implied by the second law of motion.

\section{expanding KK-space\label{kks}}
On taking Klien's method of compactification one begins with a tensor of order \ct{KK}:
\be
g_{IJ}^{(5)}  = \left( {\begin{array}{*{20}c}
   {g_{_{\mu \nu } }^{(4)} } & { +  \vee A_\mu  A_\nu } & { \vee A_\nu}  \\
   {} & { \;\vee A_\mu} & { \vee }  \\
\end{array}} \right)
\ee From equation (\ref{com}) and with an earlier work \ct{Hale}, I
choose to write a Minkowski metric of form:
\begin{eqnarray}
|(\omega,z)|^2=(\phi) c{\wedge z_1}-\omega z^2_2-\omega
z^2_3-\omega z^2_4-(\phi) c{\wedge z_5}\equiv\\{\cal I}(ct)^2-i(x)^2-j(x)^2-k(x)^2 \end{eqnarray} which is
representative of a fractal spacetime of the form $4\wedge\phi^2$.  In
tensorial terms leads to
\be
{\hat{\cal M}}=\left( {\begin{array}{*{20}c}
   i & 0 & 0 & { - 1}  \\
   0 & { - i} & 1 & 0  \\
   0 & { - i} & i & 0  \\
   { - i} & 0 & 0 & { - i}  \\
\end{array}} \right) \Rightarrow {\cal M}=\left( {\begin{array}{*{20}c}
   1 & { - 1} & 1 & { - 1}  \\
   1 & { - 1} & { - 1} & 1  \\
   { - 1} & { - 1} & { - 1} & { - 1}  \\
   { - 1} & { - 1} & 1 & 1  \\
\end{array}} \right)
\ee such that the interpretation then transverses to:
\be
{\cal{M}}^{diag}=
\left( {\begin{array}{*{20}c}
   2 & 0 & 0 & 0  \\
   0 & { - 2} & 0 & 0  \\
   0 & 0 & {2i} & 0  \\
   0 & 0 & 0 & { - 2i}  \\
\end{array}} \right) \wedge\tilde {\cal M}^{(4)}  \equiv \left( {\begin{array}{*{20}c}
   2 & 0 & 0 & 0  \\
   0 & 2 & 0 & 0  \\
   0 & 0 & { - 2i} & 0  \\
   0 & 0 & 0 & {2i}  \\
\end{array}} \right)
\label{cdim} \ee Here the time dimension is given statute through
quaternion rotations in C* space.  The superposition of
electromagnetism and gravitation can be seen within a relativistic
frame in a accordance with $\hat\eta_{IK}=diag(-2,-2,2i,-2i)$, in
the fifth coordinate this corresponds to
$\eta^{(5)}_{IK}=diag(-1,-1,-1,-1,-1)$.  In essence (\ref{cdim})
is a combination of two metrics, a similar metric was inferred in
Ref. \ct{bell} in relation to quantum gravity:
\be
d\tau^2={a\over r}\,dt^2+{a\over r}\,dr^2-dx_1^2-dx_2^2
\ee

Through some work made by Weyl \ct{Weyl} one can write a solution to EFE which corresponds to
\be
R^k_i-{1\over 2}\delta^k_iR=-{1\over 2}\nabla\psi^k_i\dots\label{weq1}\ee
which can be reduced for convenience as ${1\over
2}\nabla\psi^k_i=-T^k_i$.  Furthermore this action can be represented with advanced and retarded potentials.  When one conveniently
exchanges the $\psi$ term from equation (\ref{sup}), one is left
with the potentials
\be
\psi^k_i(x)_-=-\int{T^k_i(t-r)\over 2\pi r}dV\quad {\rm and},\quad\psi^k_i(x)_+=-\int{T^k_i(t+r)\over 2\pi r}dV
\ee

Therefore meaning that the superposition of the field is made possible
through an advanced wave.  Thus one has the compactification of a
Fouier series of form
\be
g_{IK}=\sum_n{g^{(n)}_{IK}(x^\mu)e^{inx^5/\lambda_5}}. \ee Which
under compactification yields
\be
\psi(x,x^5)={1\over\sqrt{l_p}}\sum_{n\in
2}\psi_n(\circ,x)e^{inx^5/R_5} \ee where $\circ$ represents
quarternions.   The advanced Fouier sine wave is:
\be
\psi(x,x^5)=\sqrt{{2\over\pi}}\int_0^\infty f(\circ,x)sin\: dxtdt
\ee
which undergoes the quantum transform
\be
\Psi(\circ,k,t)={1\over\hbar}\Phi(\circ,k,t)e^{ik\omega}dk\quad
and,\quad\Phi(\circ,k,t)={1\over\hbar}\Psi(\circ,k,t)e^{ik\omega}dk.
\ee

This action creates a cascade motion within the fifth coordinate
and resulting in torsion within four-dimensional spacetime.  Torsion would appear to be in form of gravitational waves
through the action
\be
\left( D^\mu D_\mu-{n^2\over R^2_5}\right)\psi_n=0. \ee Thus it is seen
that an observation will only occur in a quantum system if two
anti-symmetric $\eta^{(5)}_{IJ}$ tensors come in contact (which one might expect from the Weyl tensor).  This
wave equation suggest KK-space expands into four-dimensions,
resulting in self interaction.  Furthermore when one compares the charge $q_n=n(k/R_5)$
with the planck length, one sees the relation with the fine
structure constant.
\be
R_5={2\over\sqrt{\alpha}}l_p \ee From equation (\ref{fine}), from this it may be seen that the second law produces fine structure which in turn yields the planck length.

\section{gauge backgrounds\label{gback}}

The gravitational force is a collection of interacting forces
connected in some form by the {\em second law} (e.g. the fine structure
constant).  When one separates the properties of a given force
from the Einstein equations, its fundamental principle break
resulting in only a weak equivalence principles (which can be
interpreted as a gravitational pressure).  Thus lending a manifold
whos properties depend on the pressures applied to it by external
factors.  By the methods implied thus far it makes sense to make
use of the semi-classical approach to gravitation
$G_{\mu\nu}(\gamma)=<\psi|T_{\mu\nu}(g,\hat\phi)|\psi>$.  To begin
let us apply a gauge field of form
\be
-k(F_\nu^{\mu;\psi}-{1\over
2}\delta^{\mu\nu}_{;\psi(x)}+A_\mu^{\alpha\nu}F_{;\psi})\neq 0 \ee
which resembles a convection made seventy year ago by Einstein \ct{Ein}:
\be
G^{\mu\alpha}\:_{;\alpha}-F^{\mu\nu}\:_{;\nu}+\Lambda_\mu\:^\sigma\:_{\tau} F_{\sigma\tau}\equiv 0.
\ee
Thus it may be viewed that the above equation is the solution for flat spacetime which implies that the canonical
approach
$\gamma_{\alpha\beta}(x)=\eta_{\alpha\beta}+kh_{\alpha\beta}(x)$
should be utilized.  Such that the gauge field equation becomes:
\be
-k(F_\nu^{\mu;\psi}-{1\over 2}\delta^{;\psi
(x)}_{\mu\nu}+A_{\alpha\nu}^\mu F^{;\psi})\geq
i\hbar{\partial\psi\over\partial t}\left\{{8\pi\over\sqrt{-\tilde
g}}{\cal T}_\nu^{\mu;\psi (x)}(x)\right\} \label{emperical}\ee
where
\be
i\hbar{\partial\psi\over\partial t}=[{1\over 2m}(\hat
p-eA)^2+eV]\psi.
\ee
From this it is seen that the right of the
equation is governed by the laws of quantum mechanics giving a
pseudo unification through means of a complex gauge field.  Meaning that
the fifth coordinate is false, however through complex fields,
torsion becomes an integral part of both sides of the gauge
inequality. The stress-energy tensor can have torsion along
with electromagnetic field through the classical connection
\be
T_{\mu\nu}=(Qc^2+p)u_\mu u_\nu+pg_{\mu\nu}+{1\over c^2}(F_{\mu\alpha}F_\nu^\alpha+{1\over 4}g_{\mu\nu}F^{\mu\nu}F_{\mu\nu}).
\ee

Where torsion is given through $S_{\mu\nu\sigma}=\psi_{[\mu\nu\sigma]}$,
implying the inequality has torsion in flat spacetime; where one
may utilize the action principle \ct{tor}:
\be
\delta\int\sqrt{-g}d^4x\left({R\over k}+L\right) =0. \ee Therefore
a pseudo superposition can take place within flat spacetime,
explaining the relationship between Newtonian gravitation and
electrostatic potentials in previous sections.

\section{vacuum energy and geodesics\label{veg}}
\label{zpf} From the Dirac field $i\hbar{\partial\psi\over\partial
t}$, matter would act as a void within the QED vacuum. This would
thus cause the virtual energy ${1\over 2}\hbar\omega$ of the
quantum vacuum, to adapt a negative energy term. This process
would then act to collapse the space around it, in the presence of
$n\geq 1$ `false vacuum' mass acts on the fields to adopt a {\bf
negative energy requirement}, which violates the weak energy
condition (WEC) $T_{\mu\nu}V^{\mu}V^{\nu}\geq 0$. Here we take
this to mean a cosmological constant, such that the gauge inequality
(\ref{emperical}) becomes: \be -k(F_\nu^{\mu;\psi}-{1\over
2}\delta^{;\psi (x)}_{\mu\nu}+A_{\alpha\nu}^\mu
F^{;\psi})+\lambda\geq i\hbar{\partial\psi\over\partial
t}\left\{{8\pi\over\sqrt{-\tilde g}}{\cal T}_\nu^{\mu;\psi
(x)}(x)\right\}\ee

The cosmological constant can be given through $\Lambda=-{1\over
16\pi}F^{\mu\nu}F_{\mu\nu}$, so that the inequality suggest that
$\Delta x^\mu\Delta x^\nu\geq{1\over 2}\Lambda g^{\mu\nu}$. From this we may conclude that there exist an uncertainty within the field.  This is
impart because the vacuum can be described through:
\be
R_{\mu\nu}-{1\over 2}g_{\mu\nu}R=-\Lambda g_{\mu\nu} \ee we can
also see that this formalism closely resembles (\ref{weq1}), i.e.
Weyl's definition.  Which suggest electrostatic energy is lost
through the uncertainty which exist through the pseudo geometry
and vacuum. With
\be
F^{\mu\nu}={\partial A^\nu\over\partial x_\mu}-{\partial
A^\mu\over\partial x_\nu},
\label{empot}\ee
the geodesic for the vacuum becomes
\be
{\partial^2 A^\nu\over\partial S^2}+\Gamma_\mu^\nu\left( {\partial
x^\mu\over\partial S}\right)\left( {\partial x_\nu\over\partial
S}\right)=-{e\over mc^2}\;A_\mu\;x^\mu \ee we note that under this pseudo connection the Gamma term appears to be under torsion, through an action of $\Gamma_a^b=d\Lambda_a^b+\Lambda_a^c\wedge\Lambda_c^b$.  Thus (\ref{empot}) would appear to take the form:
\be
F_\mu^\nu={\partial A^\nu\over\partial x_\mu}-{\partial
A_\mu\over\partial x_\nu} \ee such a geodesic path is remarkably
similar to a sphere geodesic of an electron traveling through
gravitational and magnetic fields
\be
{d^2x^\mu\over ds^2}+\Gamma_\alpha^\mu{dx^\alpha\over ds}{dx^\beta\over ds}={q\over mc^2}F_\alpha^\mu{dx^\alpha\over ds}.
\ee

However, the accepted geodesic for an electromagnetic field is that of
\be
mc\left({\partial u^i\over\partial S}+\Gamma^i_{kl}
u^k\;u^l\right)={e\over c}F^{ik}\;u_k \label{emgeo} \ee From the
above equation, it can be seen that at least empirically the
vacuum electrostatic potential (the true vacuum) is responsible
for curvature of spacetime.  If one were to block the vacuum
energy as in the case of the Casimir effect, it will create an
inequality within the pseudo geometry resulting in a gravitational
pressure.  Therefore so to speak, a matter Lagrangian (false
vacuum) shields (true) vacuum (zero-point-field) energy, thus
resulting in negative energy, which may be interpreted through the
Weyl tensor as torsion.  Specifically the interaction in time
produced by the pseudo Kaluza-Klien space produces the
relationship between the vacuum states by \ct{pcos}:
\be
\phi_{out}(k,\eta)=\alpha\phi_k^++\beta\phi_k^-\label{thefalsevac}
\ee
thus acting as the advanced and retarded Fouier series seen in section (\ref{emper}).  Therefore the Riemannian tensor $R_{\mu\nu\rho\sigma}$ contracts via the Levi-Civita connection to conserve the vacuum
term, resulting in symmetric Ricci spacetime curvature, along with an antisymmetric Weyl torsion.

\section{EM vacuum and the Weyl tensor\label{emwt}}
\label{wtensor} If one takes the coefficients of the
Cosmological Constant and the Weyl tensor one has the antisymetric
field $C_{[\rho\sigma][\mu\nu]}F_{\mu\nu}F^{\mu\nu}$ (note: for
simplicity the contrivariant term $F^{\mu\nu}$, will be removed, it will be
reinstated in (\ref{cosg}))\footnote{It is exactly from this action
that we see the gauge condition envisioned by Einstein \ct{Ein}
appear in a more coherent form.}.  Which has the following form:
\begin{eqnarray}
C_{[\rho\sigma]}F_{\mu\nu}=C_{[\rho\sigma]\mu}^\mu F_{\mu\nu}=C_{[\rho\sigma][\mu\nu]}F^{\mu\nu}F_{\mu\nu}\\ =CF_{\mu\nu}=C^\mu\;_\mu F_{\mu\nu}=C_{\mu\nu}F^{\mu\nu}F_{\mu\nu}\\=C_{\rho\sigma}F^{\mu\nu}F_{\mu\nu}\label{wsym}\\C_{\rho[\sigma\mu\nu]}=0
\end{eqnarray}
from this it may be seen that the Weyl tensor is an electromagnetic
version of GR.

One may now apply a jacobi identity in order to from a pseudo
Bianchi Identity of the form $C_{\alpha\beta[\mu\nu;\lambda]}=0$.
Which can be reduced to
\be
\nabla_\lambda C_{\alpha\beta\mu\nu}+\nabla_\nu C_{\alpha\beta\lambda\mu}+\nabla_\mu C_{\alpha\beta\nu\lambda}=0
\ee
where we can contract with $F^{\alpha\mu}$
\be
\nabla_\lambda C_{\beta\nu}-\nabla_\nu C_{\beta\lambda}+\nabla_\mu
C_{\beta\nu\lambda}^\mu =0 \ee which can be contracted further
with $F^{\beta\lambda}$
\be
\nabla_\lambda C_\nu^\lambda -\nabla_\nu C+\nabla_\mu C_\nu^{\mu}=0
\ee
or
\be
\fbox{$\nabla^\mu(C_{\mu\nu}-{1\over 2}F_{\mu\nu}C)=0$}\label{pwcurve}\ee From
this we may conclude that a similar transformation will be made
with the anitsymmetric covariant term $C_{\rho\sigma}$.  When we
restore the field with equation (\ref{wsym}), we have the
following field equation:
\be
C_{[\rho\sigma]}-{1\over 4}F_{\mu\nu}+A_{\alpha\nu}^\mu C=-\Lambda g_{\mu\nu}\ee where the gauge term is assumed to be an Ansatz $A_\alpha^\mu=\eta_\alpha^{\mu\nu}\partial^\nu\ln\phi(x)$, thus equation (\ref{weq1}) is only an approximation of the above field.  To obtain field density one is left with
\be
F_\alpha^{\mu\nu}=-{\left( x^2+\rho^2\right)^2\over4\rho^2\eta_\alpha^{\mu\nu}}\label{ymdens}
\ee
 which is similar to an anitsymmetric gauge for
a Yang-Mills field see Ref. \ct{moffat}:
\be
S_{YM}=-{1\over 4}\int d^4xF^\alpha_{\mu\nu}\star F^{\alpha\mu\nu}
\ee

We recall
from section (\ref{kks}) that an observation of a gravitational
field will only occur when two antisymmetric tensor come in
contact.  Thus it is precisely the below field equation which
bridges the gap between quantum theory and GR.
\be
\left[ C_{\alpha\beta}-{1\over 4}\tilde
F_{\alpha\beta}+A_{a\beta}^\alpha C\right]_{;\beta}=0 \label{cosg}
\ee The gauge potential $A_{a\beta}^\alpha$ represents the second
component of the cosmological term $F^{\mu\nu}$ under torsion.  So that with (\ref{ymdens}), we have
\be
\left[ C_{\alpha\beta}-{1\over 4}-{4\rho^2\eta_{\alpha\beta}^a\over\left(x^2+\rho^2\right)^2}-{\left(x^2+\rho^2\right)^2\over 4\rho^2\eta_{\alpha\beta}^a}\;C\right]_{;\beta}=0\label{weyldub}
\ee
which reduces to
\be
\left[ C_{\alpha\beta}-{1\over
4}g_{\alpha\beta}+2C\right]_{;\beta}=0 \ee this solution is
parallel to the Einstein field equation, when considering
antisymmetric scalar curvatures.

Thus it is seen that the cosmological constant is the source of torsion in the antisymetric Weyl tensor.  In essence the false electromagnetic vacuum is responsible for gravitation within the Weyl-tensor.  Our new variable action is a Weyl-Hilbert action:
\be
S_{WH}=\delta\int\sqrt{-g}d^4x\left({(R+2\Lambda)\over 16\pi
G}+{\cal L}_M+{\cal L}_{YM}\right)=0 \ee resulting in an
uncertainty of the form $\Delta x^\mu\Delta x^\nu={1\over
2}|\theta^{\mu\nu}|$ (note: such an empirical uncertainty was give
in section (\ref{zpf})).  Which suggest that the Weyl tensor
should be given within a complex gauge field.  Such that the
Weyl-Hilbert action transforms to a Complex-Hilbert action of
form:
\begin{eqnarray}
S_{CH}&=&\sqrt{g}d^4x\left({(C+2\Lambda)\over 8\pi G}+{\cal
L}_M\right)\nonumber\\&&+i\left[\sqrt{g}d^4x\left({(C_{[\rho\sigma]}+2\Lambda)\over
8\pi G}+{\cal L}_M)\right)\right]\neq 0\label{comact}
\end{eqnarray}
This was suggested in section (\ref{empe}), meaning within a
quantum frame the electromagnetic and gravitational properties of
the Weyl tensor may interact through a weak superposition.
However, we also note from section (\ref{gback}) a complex solution
is only empirical, thus (\ref{comact}) is only a pseudo action.

\subsection{gravitational Lagrangians\label{glag}}
Let us form a Lagrangian for the vacuum solution $-\Lambda
g_{\mu\nu}$.  In order to describe such an action principle we
will start with the GR Lagrangian for matter in the form:
\be
S_M=\int\sqrt{g}d^4x(g^{\mu\nu}\partial_\mu\phi\partial_\nu\phi+\dots)
\ee from our approach thus far we would like to consider
perturbations from the vacuum such that:
\be
\delta_{metric}S_{EM}=-\int\sqrt{g}d^4x\Lambda g_{\alpha\beta}\delta \tilde F^{\alpha\beta}
\ee
which can simply be given by
\be
\Lambda g_{\alpha\beta}:={1\over\sqrt{g}}{\delta\over\delta \tilde F^{\alpha\beta}}S_{EM}.
\ee
From section (\ref{wtensor}) we now make the modification
\be
\Lambda g_{\alpha\beta}:={1\over\sqrt{-g}}{\delta{\cal
L}_{EH}\over\delta \tilde
F^{\alpha\beta}}S_{WH}=([C_{\rho\sigma}-{1\over
4}g_{\alpha\beta}+2C]_{;\beta}). \ee It is seen from this
Lagrangian that the cosmological constant in the EFE, is in fact
an electromagnetic Weyl tensor.

On taking the conditions $T_{\mu\nu}(x)=0\Rightarrow
R_{\mu\nu\rho\sigma}=C_{\mu\nu\rho\sigma}(x)$, in the absence of a
matter Lagrangian torsion is carried by the symmetric Weyl tensor
or generated from the electromagnetic vacuum $-\Lambda
g_{\mu\nu}$. Resulting in an uncertainty of the form $\Delta
x^\mu\Delta x^\nu={1\over 2}|\theta^{\mu\nu/
i\alpha}h_i^a\partial^\mu_\alpha h_a^\nu|$. With this one has
torsion resulting in a teleparallel description of gravitation in
flat spacetime. Thus the cosmological constant is a perturbation
within the curvature connection made possible through virtual
particles (i.e. the false vacuum).

\subsection{teleparallel geometry\label{tpg}}
The Cartan torsion connection is given by:
\be
T^\sigma\;_{\mu\nu}=\Gamma^\sigma\;_{\nu\mu}-\Gamma^\sigma\;_{\mu\nu},
\ee
with tetrad form
\be
\Gamma^\rho\;_{\mu\nu}=h_a\;^\rho\partial_\nu h^a\;_\mu
\ee
Thus the vacuum-energy tensor has torsion of the order
\be
-\Lambda{\cal G}^\beta_\alpha=h^a\;_\beta\left( -{1\over\sqrt{-g}}
{\delta{\cal L}_{EH}\over\delta h^a\;_{\alpha}}\right)
\ee
Hence the gravitational field equation from Weyl torsion is seen through
\be
\left[ C_{\alpha\beta}-{1\over 4}g_{\alpha\beta}
+2C\right]_{;\beta}=-\Lambda {\cal
G}^\beta_\alpha \ee Thus the Weyl torsion tensor within GR can be
given through the metric
\be
g_\alpha=\eta_{\alpha\beta}+\lambda{\cal G}_\alpha^\beta
\ee
such that within the EFE the Cosmological Constant takes the form
\be
R_{\alpha\beta}-{1\over 2}g_{\alpha\beta}R=-8\pi
GT_{\alpha\beta}+\lambda g_\alpha \ee It is known that the
electromagnetic field and the stress-energy tensor can feel
torsion through the action
\be
{\cal T}_{\mu\nu}=-{2\over\sqrt{-g}}{\delta{\cal
L}_{EM}\over\delta g^{\mu\nu}}={1\over 4}\left[ F_\mu^\rho
F_{\mu\rho}-{1\over
4}g_{\mu\nu}F_{\rho\sigma}F^{\rho\sigma}\right] \ee Therefore the
torsion of ${\cal G}_\alpha$ can be carried not only through
the stress energy tensor but onto G itself.

Covariant Maxwell's potentials can be given through torsion by
\be
F^*:=-\omega_\nu\wedge\theta^\nu=A_\mu^\nu\wedge\omega_\nu\wedge\omega^\mu
\ee with current potentials
\be
J^*:=dF^*=d\theta^\mu\wedge\omega_\mu-\theta^\mu\wedge d\omega_\mu
\ee
The Weyl tensor can represent such a current through
\be
\nabla^\mu C_{\mu\nu\rho\sigma}=J_{\nu\rho\sigma}
\ee
with
\be
J_{\nu\rho\sigma}=k{n-3\over n-2}\left[\nabla_\rho
T_{\nu\sigma}-\nabla_\sigma T_{\nu\rho}-{1\over
n-1}\left[\nabla_\rho T_\lambda^\lambda
g_{\nu\sigma}-\nabla_\sigma T^\lambda_\lambda
g_{\nu\rho}\right]\right] \ee Thus it is seen that the uncertainty
by the relation $\Delta x^\mu\Delta x^\nu\geq{1\over
2}|\theta^{\mu\nu}|$ causes the Weyl tensor to carry a
cosmological charge current.  Thereby the final form of the Weyl
gravitational field is
\be
\fbox{$\left[ C_{\alpha\beta}-{1\over 4}g_{\alpha\beta}
+2C\right]_{;\beta}=-J_\beta\Lambda {\cal
G}^\alpha$} \ee The right side of the above equation can be
written in the form $-J\Lambda{\cal G}_\alpha^\beta$.  From our
pseudo Weyl Curvature (\ref{pwcurve}), the relationship between
charges is given by:
\begin{eqnarray}
\nabla^\mu C_{\mu\nu}+\left[\nabla_\rho T_{\nu\sigma}-\nabla_\sigma T_{\nu\rho}\dots\right]\equiv J_{\nu [\rho\sigma]}\longrightarrow\\\nabla^\mu F_{\mu\nu [\rho\sigma]}=J_{\nu}-\left[\nabla_\rho T_{\nu\sigma}-\nabla_\sigma T_{\nu\rho}\dots\right]
\end{eqnarray}
This approximation can be given through:
\be
F_{\mu\nu}={\partial\Phi_\nu\over\partial x^\mu}-{\partial\Phi_\mu\over x^\nu}+C_{\mu\nu}^{\alpha\beta}\Phi_\alpha\Phi_\beta
\ee
which is made possible through a Ricci symmetric tensor of form
\be
{\partial\Lambda_\mu^\sigma\over\partial
x^\nu}=\eta_{\mu\nu\sigma}\Phi_\nu\Lambda_\mu^\sigma \ee which in
a constant field is given by $R_{\mu\nu}=\Lambda_\mu^\sigma
F_{\sigma\nu}$, such a field can transpose to the relation
\be
F_{\mu\nu}=\partial_\mu\Phi_\nu-\partial_\nu\Phi_\mu
\ee
such a field describes two opposed electromagnetic fields through the connection $\Gamma_{\mu\nu}^\sigma=\Lambda_\mu^\sigma\Phi_\nu$.  Which has the geodesic relation
\be
{d^2x^\mu\over ds^2}+\left(\Phi_\sigma{dx^\sigma\over ds}\right)\Lambda_\nu^\mu\;{dx^\mu\over ds}=0
\ee
the above geodesic also has the form
\be
\Phi_\sigma{\partial\Lambda_\mu^\sigma\over\partial x^\nu}=C_{\nu\sigma}^{\alpha\beta}\Phi_\alpha\Phi_\beta\Lambda_\nu^\sigma
\ee
this term is thus antisymmetric and yields
\be
R_{\mu\nu}=\left({\partial\Phi_\nu\over\partial x^\sigma}-{\partial\Phi_\sigma\over x^\nu}+C_{\nu\sigma}^{\alpha\beta}\Phi_\alpha\Phi_\beta\right)\Lambda_\mu^\sigma=F_{\sigma\nu}\Lambda_\mu^\sigma
\ee
One should also note that a four vector line element under this prescription is given by
\be
u_q={\Phi_{;q}\over\sqrt{-\Phi_;\Phi^q}}
\ee
Finally a like wise connection is given through
\begin{eqnarray}
C_{\rho\sigma\mu\nu}= R_{\rho\sigma\mu\nu}-\biggl({2\over (n-2)}g_\rho [_\mu R_\nu]_\sigma-g_\sigma[_\mu R_\nu]_\rho\nonumber\\+{2\over(n-1)(n-2)}Rg_\rho[_\mu g_\nu]_\sigma\biggr)
\end{eqnarray}
Thus allowing a metric to remain conformaly invariant through the
rescaling operation $g_{\mu\nu}(x)\rightarrow
e^{f(x)}g_{\mu\nu}(x)$.  This would make the torsion term appear
to disappear within classical GR by means of Ricci curvature, or
through the Einstein-Hilbert action $S=\int\sqrt{g}dx^4R$.
However, the cosmological charge current term would still remain
connected to the stress-energy tensor.  This result may be
obtained through the Tucker-Wang action
\be
S=\int\lambda^2R\star 1 \ee where $R\star 1=R_b^a\wedge(e_a\wedge
e_b)$, is a scalar torsion  corresponding to
$T^a=de^a\Lambda_b^a\wedge e^b$.  With this one can have an
action principle which resembles a Brans-Dicke space by:
\be
S=\delta\int\lambda^2\left({R\star 1\over 16\pi G}+{\cal
L}_{YM}\right)=0\label{weylact} \ee Thus we have a action
corresponding to a Cosmological Constant, without $\Lambda$. This
is made possible because we have been considering a Weyl action,
empirically given by:
\be
S_W=-\alpha\int
C_{\lambda\mu\nu\kappa}C^{\lambda\mu\nu\kappa}\sqrt{-g}dx^4. \ee
Since the Cosmological coefficient is included in the Weyl tensor
we have been considering, it vanishes under a Cosmological model.
Therefore the action for Weyl gravitation is not governed by the
pseudo action (\ref{comact}), but by (\ref{weylact}).

\section{The vacuum and the meaning of G\label{vmg}}
An alternative interpretation of mass was assumed by de Broglie by
the Einstein-de Broglie equation:
\be
\hbar\omega_C=m_0c^2. \ee This formalism has be restated recently
by Haisch and Rueda \ct{haisch} as a possible explanation for the origin of
inertial mass.  Where C is given by the Compton wavelength
$\lambda_C=h/mc$, thereby asserting the origin of inertia through
the Compton wavelength.  If we take the equivalence principle by
heart then, one must assume that gravitational inertial would
arise through a similar action.  From Einstein-Cartan geometry we
can assume that this field would be given through torsion.
Specifically we will assume an equivalence of order
$\Lambda_C=2\pi G^{-1}$ (see \ref{comp} for details),
this is validated by the quantization $\lambda_C/2\pi$.  From this
we see that G is the inverse charge of an electron's Compton
wavelength.  In terms of EFE we have
\be
R_{\alpha\beta}-{1\over 2}g_{\alpha\beta}R=-{8\pi\left\langle
G_Q\right\rangle\over c^2}T_{\alpha\beta}+\left\langle\lambda_
Q\right\rangle{\cal G}_\alpha \ee in essence it appears that this
is simply a post Einsteinian semi-classical correction to the
field equations (with teleparallel Weyl torsion acting as a gauge
background) .

From our supposed relation, one would have an identity of form
$\lambda G=I$.  Therefore the above relation transverses to
\be
R_{\mu\nu}-{1\over 2}g_{\mu\nu}R+\lambda -G=-8\pi IT_{\mu\nu}
\label{efei} \ee Let us now suppose that the identity is
equivalent to the de Broglie wavelength $I=h/p$.  Thus:
\be
\left\langle G_Q\right\rangle\equiv{h/p\over h/mc}={\lambda_d\over\lambda_C} \ee where
$\lambda_d$ is the de Broglie wavelength given by $\lambda_d=h/p$.  From Compton scattering we may assume that gravitational waves can pass through particles as though they were a wave.  When we view the geodesic (\ref{emgeo}) one may rewrite the identity in eq. (\ref{efei}), such that
$I=1/\sqrt{g_{kk}}$.  This results from a geodesic of order
\ct{loupwarp}:
\be
{1\over\sqrt{g_{kk}}}\left[{\partial^2 u^i\over\partial
S^2}+\Gamma^i_{kl}\left({\partial x^k\over\partial
S}\right)\left({\partial x^l\over\partial S}\right)\right]={e\over
mc^2}F^{ik} \ee thus it is viewed from this that the term $e/mc^2$
is nearly a classical approximation of the Compton wavelength
$\lambda_C=h/mc$.  We see this when we compare the energy sources
to that of the classical fine structure constant
\be
\sigma_{rad}\sim{Z^2\over 137}\left({e^2\over
m_0c^2}\right)^2\;cm^2/nucules \ee which is given by an electrons
radius $r=e^2/m_0c^2=2.818\times 10^{-13}cm$.  Where through the
quantum correction one has $2\pi e^2/hc={1\over 137}$.  Further the
inversion by the Cosmological Constant\footnote{This prescription
of the Cosmological Constant as the Compton wavelength
$\lambda_C$, may have bearing on modern cosmological theories.
For example it has been proposed by recent observations from type
IA supernova, there may be something causing the universe to
accelerate its expansion.  Under this scenario, the Universe would
be coasting from the initial 'big bang,' however through a
cosmological Compton scattering this effect would appear to
increase, thus giving the allusion of an `accelerating' universe.}
yields the Gravitational Constant through the identity
$1/\sqrt{g_{kk}}$, or specifically through the geodesic coordinate
$u_k=\partial S/\partial x^k$.  Thus it is seen that the
gravitational constant, fine structure, and the second law of
motion appear to arrive from quantum charges!

\subsection{standing waves and the fifth coordinate\label{swf}}
We note that a superposition of a sinusoidial wave yields a standing
wave of the form $p=2a\sin (2\pi x/\lambda )\cos (2\pi vt/\lambda
)$.  With this the standing wave of the gravitational constant would be given through:
\be
p_G-\lambda_0=2a{\cos(2\pi vt/\lambda_d)\over \sin(2\pi
x/\lambda_C)}=\cot\;2a{vt\lambda_d\over x\lambda_C}\label{gsup}
\ee Thus G is the inverse ratio between the superposition of de
Broglie and Compton wavelengths\footnote{The idea of a quantum
connection to the Gravitational Constant and the Cosmological
constant is not new idea, and neither is a superposition relation
see Ref.\ct{pcos}.}.  This standing wave can be seen as potential
barrier, which results in the interaction of advanced and retarded
potentials\footnote{An alternative to this interpretaion arises
through Quantum Mechanics, from SR an electromagnetic field at k
reading locally as electric may read as a magnetic field in the
frame k'.  Through the action of the Weyl tensor the electric and
magnetic terms may become superimposed, thus initially one has a
superposition of form $\psi(x)=E(x)+H(x)$, which doesn't take on
its SR form until the wave function has been canceled.}.  This
action results in the violation of the WEC, i.e. results in a
false vacuum and Weyl gravitation.

From this we understand that a gravitational constant is an
inverse charge of a particles Compton wavelength.  Meaning that
each particle has its own local isotropic gravitational field,
which is induced by mass and acceleration.  This also leads to a startling
corollary under relativistic velocities the charge of spacetime Q
would be altered.  In such a situation the Compton charge would be
altered by
\be
{\cal L}=mc^2\sqrt{1-{v^2\over c^2}}-m\Phi c-q\Phi E+q\vec v\vec
A, \ee causing an inverse relation in the G (the consequences of
such an effect is briefly mentioned in \ref{lmass}).  Since the
electromagnetic force has an inverse relationship squared to
infinity, thus is the gravitational field.  Forces such as the
Yang-Mills field are confined within the Coloumb barrier, thus
allowing the gravitational field at a first approximation to adopt
the Newton's Gravitational constant $G_N$.  When compared to
classical gravitation one has the field $\nabla\varphi=4\pi
[2a{vt\lambda_d\over x\lambda_C}]_\rho$.  From this a four-vector
is required, thus gravitation is a charge in spacetime!

Through Ref. \ct{tkk} we see our prior assumed equivalence with the Compton wavelength pops up again through the five-dimensional action:
\be
\Psi (x^\mu,\;x^5)=\exp\left[
ik{2\pi\over\lambda_C}x^5\right]\;\psi (x^\mu) \ee where it is
interpreted that the gravitational force arises through torsion.
While electromagnetism is derived through the fifth
gauge-component of the torsion tensor (which has been shown to be
false in previous sections).  This is a pseudo {\bf complex
interpretation} produced by (\ref{gsup}) and (\ref{thefalsevac}),
thereby giving the allusion to a `fifth-coordinate,' through a
superposition mechanism.

\section{discussion\label{dis}}
This analysis of a gravitational background space reveals the
following subtle quantum aspects of the gravitational field.  The
dual interpretation of a causal trajectory in the Feynman school,
is responsible for the appearance of a pseudo `fifth coordinate.'
Thus causing true vacuum energy to translate into false vacuum energy converting the potential virtual energy into kinetic energy.  This results in torsion within the background space, which acts to conserve the negative energy created by the false vacuum. Torsion then acts to produce a gravitational metric by means of a quantum charge, where by the {\em equivalence principle} the {\em second law} becomes valid for a classical body.  Secondarly torsion alters the de Broglie wavelength which causes electrostatic potentials to lower, acting as a relativistic gravitational field.

This analysis showed the importance of the often neglected Weyl
component of Riemannian geometry.  It is the antisymmetric Weyl
tensor acting along with an Einstein-Cartan geometry that
is responsible for the gravitational constant.  Specifically
pertaining to an electrons Compton wavelength for long range
gravitation.  However, for field of varying charge one would
expect the gravitational constant and the cosmological constants
to accept different values, thus gravitation in its true form
would carry more than the background electromagnetic vacuum, and
gravitation would be expected to have ranges limited to there
local fields\footnote{For example, nuclear fields do not
correspond to the inverse square relationship $1/r^2$, thus
gravitation would be expected to behave fundamentally different
here. Since the gravitational force is one of a collective nature,
all vacuum field sources should be included within such a potential
formalism.}.

\section*{Acknowledgement}
I would like to thank Fernando Loup for his correspondence on the Cosmological Constant, and for the support of this work.

\appendix

\section{The Compton wavelength and `superluminal' shifts}\label{comp}
The relationship between the gravitational constant G, and the
Compton wavelength can be seen, when we except the value of G to
be of order \ct{G}:
\be
G=(6.74215\pm 0.000092)\times 10^{-11}m^3 kg^{-1}s^{-2}. \ee We
can now compare that to the Compton wavelength\footnote{Data on
the Compton wavelength comes from \ct{compton}.} of an electron
given by $\lambda_C=h/m_ec=2.426\times 10^{-12}\; m$.  However
from field density relationship shown in (\ref{weyldub}), we are
left with the crude relation $\lambda_C=2\pi G^{-1}$, we must
consider the quantized wave form:
\be
{h\cdot\lambda_C\over 2\pi}=386.1592642(28)\times 10^{-15}\;m
\ee
Thus the inverse of G is that of
\be
2\pi G^{-1}=0.02361=\lambda_C\times 10^{-1} \ee Such that it is
seen that $2\pi G$ is the inverse of the Compton wavelength.
Acceleration will altered the `charge' or Compton wavelength
such that
gravitational constant(s) would be altered upon relativistic velocities. This
could very well explain the propagation of observed superluminal
jets emanating from Active Galactic Nuclei (AGN).

Since classically electromagnetic waves propagate via the relation
$C=\lambda\nu$, we expect a gravitational shift from
\be
\Delta v\approx v_i(G_aM)\left({1\over r_i}+{1\over r_f}\right).
\ee thus when viewed parallel to the direction of travel one is
left with a blue shift by
\be
C_0\left( 1+{\Phi_g\over c^2}\right)=\lambda_0\left(\nu_0\left( 1+{\Phi_g\over c^2}\right)\right).\label{gspeed}
\ee
Where
\be
\Phi_g=-{G_Qm_0\over\sqrt{x^2+(y^2+z^2)(1-v^2/c^2)}}
\ee
the wavelength would appear to be altered by
\be
{\cal L}_\lambda=\lambda'\sqrt{1-{v^2\over c^2}}\label{llag}\ee
this is because (\ref{gspeed}), makes it appear that the
wavelength $\lambda$ is increasing while it is really of function
of $C_0$ and $\nu_0$.  This would appear to yield superluminal
travel, such a result is in accordance with \ct{blazar}.  This
effect does not seem to be limited to AGN either, a superluminal
source was also detected near SN1987A, see \ct{SN}.

\section{origin for inertia and mass increase?\label{lmass}}
From the previous section we have seen that the gravitational constant could be considered as the inverse of the Compton wavelength.  From (\ref{gspeed}), we now may consider an inverse of the quantum energy $E=hv$ and the classical wavelength $C=\lambda\nu$:
\be
\lambda_d={E\over C_0\left( 1+{\Phi_g\over
c^2}\right)}={h\nu_0\over\lambda \left(\nu_0\left( 1+{\Phi_g\over
c^2}\right)\right)}\rightarrow{h\over\lambda \left(\nu_g\left(
1+{\Phi_g\over c^2}\right)\right)}\equiv I\label{dbrogl}\ee such
that acceleration yields a mass increase through the de Borglie
relation
\be
p_0\left( 1+{\Phi_g\over
c^2}\right)=mv_0=\lambda_C\left(\nu_0\left( 1+{\Phi_g\over
c^2}\right)\right) \ee thus mass increase is governed by the
action seen in (\ref{llag}).  This is thus a verification of the
equivalence principle, i.e. inertial and gravitational masses are
equivalent!  From (\ref{dbrogl}) we can now consider a Lagrangian
of form:
\begin{eqnarray}
\lambda_d'&=&m_0c^2-q\Phi'{E\over C_0\left( 1+{\Phi_g\over c^2}\right)}\nonumber\\&=&m_0c^2-q{\left(\Phi {E\over C_0\left( 1+{\Phi_g\over c^2}\right)}-v\cdot A/c\right)\over\sqrt{1-{v^2\over c^2}}}
\end{eqnarray}
this interpretation runs parallel with \ct{krough}.  However, this
work diverges with equation (\ref{gspeed}), thus it is $\lambda$
which creates observable gravitational effects and not $\nu$.
With this in mind one can have an action of
\be
\Delta S={m_0c^2\over h}\sqrt{1-{v^2\over c^2}}+ {\lambda_C\over
h}\int\Phi_gdt, \ee therefore the appearance of inertia only
appears for particles with a corresponding Compton wavelength
through the action
\be
\Delta S={\lambda_C\over h}\int(\Phi_g-v\cdot A/c).
\ee

\section{gravitation within the QED vacuum}\label{gqd}
It is known that particles such as the proton have a value of
$\lambda_c=h/m_pc=1.321\dots\times 10^{-15}\;m$, for the Compton
wavelength.  Meaning that Gravitation is not a force directed by
one term, but all terms of vacuum.  Thus it maybe seen that
gravitation within an nucleus behaves quite differently than the
Newtonian prescription.  We assume from elementary data that a
`nuclear' gravitational field would be confined to the nucleus,
not manifesting its effects in the global sense.  However, for the
early universe, one may have a spacetime with quite different
cosmological constant(s) then the ones observed today, possibly
giving new justification for inflation theory.  Lastly in comparison to Appendix \ref{comp}, a beam of protons being accelerated from an AGN source would result in another prediction.  The proton/electron `acceleration' rates, and for any particle in general is directly
proportional to their Compton wavelengths.

\section{implications for the planck length\label{impl}}
It is believed that the planck length $l_p=(Gh/c^3)^{1/2}$, is the
fundamental cut off point for the gravitational field.  Two
problems arise with this work 1) the planck length is determined
by the Compton wavelength of the mass in question.  2)  the
gravitational constant and thus the planck length are altered upon
acceleration.  The first problem is not a problem it is simply a
modification required by the theory, and for the large scale
universe this result is negligible under a first approximation.
The second problem is still a problem, however in an earlier work
\ct{Hale}, I modified the planck length with disconcern.  However,
with that work in mind problem two is easily solved and is given
by
\be
l_p=(G_Ch/m_{p_0}c^3)^{1/2}\cdot\psi.\label{npl} \ee Where $G_C$
is the gravitational constant given by the Compton wavelength, for
an electron this becomes Newton's gravitational constant $G_N$.
And  $ m_{p_0}$ is the rest momentum of the mass in question,
which is given by
\be
m_{p_0}=\mp (pc/c^{-2}). \ee This definition is given by the
relativistic wave equation $E=\pm(pc+m_0c^2)$.  From (\ref{npl})
the gravitational constant can also be considered in the form
\be
G_C={2\pi c^3l_p^2\over h}=2\pi c^3(G_C^2
h/m_{p_0}^2c^6)^{1/2}=2\pi {G_C{1\over 2}hc^3\over
m_{p_0}c^3}=2\pi{G_C{1\over 2}h\over m_{p_0}}\label{qg2} \ee With
this we see a gravitational uncertainty through $\Delta x\Delta
p\geq {1\over 2}Gh$.  Finally after quantization of (\ref{qg2}) we
have a pure quantum charge, i.e $G_Ch$, thus gravitation carries
the uncertainty of the Compton wavelength.

\bibliographystyle{plain}

\end{document}